# Efficient and Robust Spatial-to-Fiber Coupling for Multimode Quantum Networks via Cascaded Adaptive Feedback Control

Ya Li, WanRu Wang, Weizhe Qiao, Qizhou Wu, Changqing Niu, Xiaolong Zou, Youxing Chen, Xin Guo

*Abstract*—Duan-Lukin-Cirac-Zoller (DLCZ)-based multimode quantum networks rely on efficient spatial-to-fiber coupling, yet environmental perturbations compromise this performance. We develop a cascaded adaptive feedback control system integrated into the quantum entanglement source preparation path. Leveraging a power-feedback hill-climbing algorithm, it dynamically regulates piezoelectric-actuated mirrors to achieve autonomous multi-dimensional beam alignment. Experiments show it rapidly boosts single-mode fiber (SMF) coupling efficiency to over 70% within 20 seconds and entering the most efficient and stable transmission state after 75 seconds. Importantly, it enhances the stability of the atom-photon interface—critical for quantum light-matter interactions—providing a practical framework for efficient, robust spatial light transmission in scalable quantum networks.

*Index Terms*—Adaptive optics, Quantum networks, Fiber coupling, Feedback control, Piezoelectric actuation.

## I. INTRODUCTION

QUANTUM communication enables theoretically unbreakable secure transmission via light, where the multimode multiplexed Duan-Lukin-Cirac-Zoller (DLCZ) protocol stands as a cornerstone for scalable long-distance networks[1]. It generates entangled Stokes photon-spin wave pairs through spontaneous Raman scattering—serving as quantum repeaters—while leveraging inherent simplicity and compatibility with existing fiber infrastructure [2–5]. However, scaling such systems faces critical bottlenecks: efficient multimode entanglement preparation, storage, and readout demand strict phase-matching [6-9], necessitating sub-wavelength alignment stability between writing/readout lasers and atomic ensembles [10,11]. Although multiplexing advances (e.g., 225-mode 2D arrays, time-bin multiplexing [12,13]) enhance capacity, escalating mode counts exacerbate complexity in multi-node cascaded architectures, where inter-node optical paths form fragile networks. Minute perturbations—beam displacement, angular jitter, wavefront aberration — degrade single-mode fiber (SMF) coupling efficiency and amplify instability across cascades, directly undermining reliability [14–16]. This is pivotal: while a 1% gain in multimode readout efficiency boosts entanglement distribution rates by 7–18% [9,13], achieving stable spatial-to-SMF coupling remains the critical barrier.

Adaptive optics (AO)[17-19] offers promising solutions: nonlinear acousto-optic deflector (AOD) drives enable 30-kHz cylindrical aberration correction (λ/35 accuracy @800 nm) [20]; integrated tip-tilt sensors leverage auxiliary-fiber feedback for real-time misalignment detection [21]; laser spiral scanning suppresses dynamic disturbances [22]; stochastic parallel gradient descent (SPGD) algorithms achieve >52% SMF efficiency [23]; multi-plane light conversion (MPLC) boosts coupling from <0.1% to 58.61% [24]; and estimation-based SPGD (ESPGD) enhances jitter-correction bandwidth [25].

DLCZ based multimode quantum networks achieve large-scale communication through multidimensional multiplexing, yet their performance is limited by dual bottlenecks in spatial-to-fiber coupling: first, the parallel transmission of multiple quantum states requires strict maintenance of mode orthogonality—coupling deviations easily trigger crosstalk, resulting in the loss of the capacity advantage of multimode systems. second, loss accumulation in multi-node cascades and the high sensitivity of quantum states to perturbations further restrict fidelity and stability. existing solutions (e.g., MPLC focusing on wavefront aberration correction, SPGD emphasizing general perturbation optimization) fail to meet the "rapid response" and "long-term stability" requirements of multimode scenarios. to address this, this study proposes a cascaded adaptive feedback control system: by simulating the average light field of multi-modes via an auxiliary optical path and dynamically adjusting mirrors using a power-feedback hill-climbing algorithm, the system not only eliminates the need for independent optimization of individual modes but also rapidly boosts the coupling efficiency to over 70% within 20 seconds and enters the most efficient and stable transmission state after 75 seconds. this simultaneously resolves the "timeliness" and "stability" dilemmas in multimode coupling, providing key

This work was supported in part by the National Natural Science Foundation of China (U23A20636, 62204232) , Supported by Fundamental Research Program of Shanxi Province(202303021222109) and the Natural Science Foundation of Shanxi Province (20210302124189, 202303021212208, 202403021222162). Research Project Supported by Shanxi Scholarship Council of China (20210038).

Ya Li, WanRu Wang, Qizhou Wu, Changqing Niu, Xiaolong Zou,Youxing Chen and Xin Guo are with School of Information and Communication Engineering, North University of China, Taiyuan 030051, China, and also with Shanxi Province Key Laboratory of Intelligent Detection Technology and Equipment, North University of China, Shanxi Taiyuan 030051, China ; (guoxin2019@nuc.edu.cn).

Weizhe Qiao is with Shanxi Dazhong electronic information Industry Group Co., LTD., Taiyuan 030024, China;

technical support for the efficient operation of multimode quantum networks.

## II. SPATIAL LIGHT-FIBER ANGULAR OFFSET MODEL

In the optical path system for preparing quantum entanglement sources, photons are transmitted via optical fibers. In experiments, efficient coupling between optical fibers and free-space optical paths is achieved using fiber collimators, thereby enabling photon input and reception. Specifically, at the transmitting end, the laser output from a single-mode fiber is adjusted by a collimator to generate an approximately parallel Gaussian beam. At the receiving end, the collimator couples the received beam back into a single-mode fiber of the same type. The energy coupling efficiency between the two ends is determined by the overlap degree of the two output Gaussian light fields.

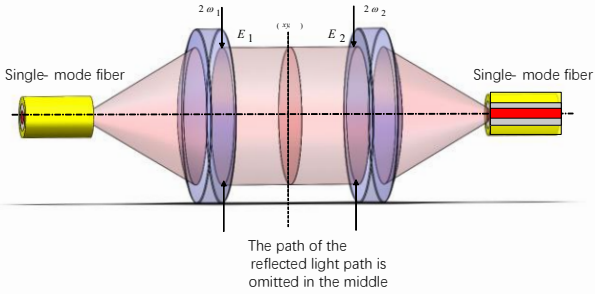

**Fig. 1.** Schematic diagram of output laser coupled to single-mode fiber

As shown in Fig. 1, the laser output from the fiber collimator is approximated as a collimated beam. Its propagation path within the free-space optical field is omitted in this figure, which only illustrates the optical field distributions of the output and coupling components under ideal conditions. Specifically, the Gaussian beams emitted from the two fiber collimators are characterized by their respective optical field distributions at the beam waist positions, denoted as $E_1(x, y)$ and $E_2(x, y)$ respectively, where:

$$E_1(x, y) = A_1 \exp\left(\frac{-(x^2 + y^2)}{\omega_1^2}\right) \quad (1)$$

$$E_2(x, y) = A_2 \exp\left(\frac{-(x^2 + y^2)}{\omega_2^2}\right) \quad (2)$$

Where $A_1$ and $A_2$ represent amplitude constants, $x$ and $y$ represent the transverse spatial coordinates describing the spot position in the cross-sectional plane, and $\omega_1$ and $\omega_2$ refer to the beam waist radii of the Gaussian beams output from the two fibers (equivalent to the output spot radii of the fiber collimators).

Based on the mode-field coupling theory, their energy coupling efficiency $T$ is expressed by equation (3):

$$T = \frac{\left|\iint E_1 \cdot E_2^* dxdy\right|^2}{\iint |E_1|^2 dxdy \cdot \iint |E_2|^2 dxdy} \quad (3)$$

Where $E_2^*$ is the complex conjugate of $E_2$, and this formulation accounts for the "phase-matched energy overlap" calculation.

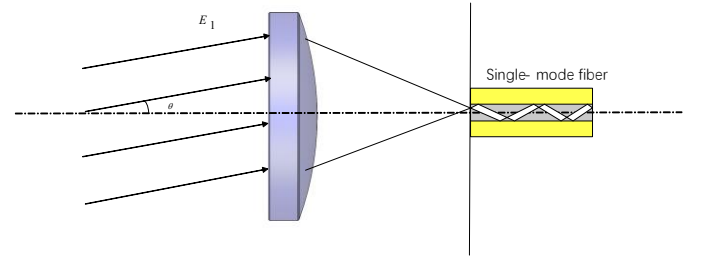

**Fig. 2.** Schematic of the angular deviation of the laser incident on the fiber collimator

Identical collimators were used in the experiment, hence $\omega_1 = \omega_2$. Under ideal conditions with no lateral offset, no angular deviation, and perfect mode-field matching, T = 1. In practice, however, the actual coupling efficiency typically ranges from 70% to 80% due to diffraction, optical path losses, optical component losses, and imperfect mode-field matching; this is defined as the intrinsic baseline efficiency, denoted as $T_{\text{base}}$. Additionally, as shown in Fig. 2, the collimated beam from the fiber collimator has a relatively large spot radius, making it highly sensitive to angular deviations. The influence of angular deviations on coupling efficiency is as follows:

$$\eta = T_{\text{base}} \exp\left(-\left(\frac{\pi \omega_1 \theta}{\lambda}\right)^2\right) \quad (4)$$

As shown in Fig. 3, the two curves depict the relationship between coupling efficiency and angular deviation under ideal conditions and in the presence of intrinsic loss, respectively. Both coupling efficiency curves follow a Gaussian function decay law. When the angular deviation is small, changes in the optical field overlap region have a weak influence on the "tail" of the Gaussian distribution, corresponding to a slower efficiency decay. When the angular deviation is close to the Rayleigh angle, the curve curvature is maximum, corresponding to a faster efficiency decay. When the angular deviation is much larger than the Rayleigh angle, the optical fields are almost completely mismatched, and the corresponding efficiency decay approaches the lower limit of the baseline loss.

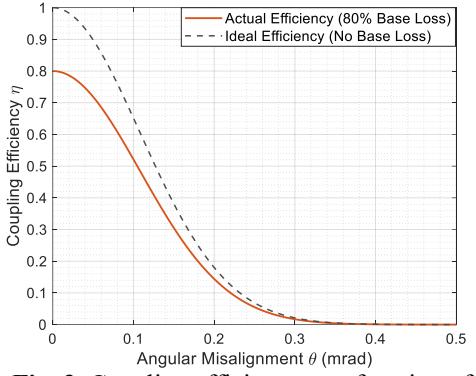

**Fig. 3.** Coupling efficiency as a function of angular offset

### III. EXPERIMENTAL SETUP AND ADAPTIVE PROGRAM DESIGN

*A. Space optical-single mode fiber cascade automatic control device*

The cascade automatic control device designed in this paper simulates the photon collection process via an auxiliary optical path and optimizes the photon-receiving path by adjusting the auxiliary light receiving efficiency through a feedback system, as shown in Fig. 4.

The atomic ensemble (rubidium atomic cluster) used in the experiment is composed of cold $^{87}$Rb atoms, which serve as the medium for generating photons with adjustable wave packets. Initially, the $^{87}$Rb atomic ensemble is in the ground state $|g\rangle$. A writing pulse induces a spontaneous Raman transition $|g\rangle \to |s\rangle$, which generates a Stokes photon transition $|e\rangle \to |s\rangle$ while storing spin waves. A reading pulse $|s\rangle \to |e\rangle$ then triggers excitation and emits anti-Stokes photons, completing the readout of photon information.

The device primarily comprises four interconnected components with clear functional division and coordinated operation to enable efficient monitoring and control of quantum photon transmission: Part 1 is the atomic-laser coupling feedback module (optical path: red lines), including a laser, FRM A, FRM B, FRM C, a photodetector, and an oscilloscope; it takes the absorption dip spectrum displayed on the oscilloscope as the core observation target, forms a closed-loop feedback system with FRM A, FRM B, and FRM C, and adjusts the incident laser angle based on the spectral depth and width (which reflect atomic-laser coupling strength) to ensure the laser interacts with the atomic cluster at the optimal angle for maximum photon excitation rate. Part 2 is the multifunctional optical path & actuation control module (optical path: arrows marked; photon propagation: green lines; auxiliary light propagation: purple lines), integrating a laser, FRM 12, FRM 34, a fiber collimator, a beam splitter, a power meter, a piezoelectric linear actuator, a host computer, FRM D, FRM E, and a single-photon detector; it constructs the main transmission path for photons and auxiliary light, while the host computer controls the piezoelectric linear actuator to adjust the tilt of FRM 12 and FRM 34 in real time, laying the foundation for precise correction of light spot angular offset. Part 3 is the signal separation & real-time monitoring module, which uses a dichroic mirror to address the frequency band difference between auxiliary light and photons—transmitting auxiliary light into the power meter for real-time monitoring of transmission efficiency fluctuations and reflecting photons into the single-photon detector to avoid signal interference and enable simultaneous data acquisition. Part 4 is the photon collection area, which leverages the linear growth relationship between photon excitation yield and writing laser power to indirectly monitor coupling efficiency via auxiliary light power, based on the principle that sufficiently high auxiliary light power corresponds to sufficiently high photon reception efficiency, simplifying monitoring while ensuring control accuracy.

During the experiment, the system collects real-time optical power information via the power meter and transmits the data to the host computer rapidly. The host computer is connected to the driver and controls the step count and step size of the actuator according to feedback signals and control algorithms. This adjusts the left-right tilt of frame adjustment knobs 1, 2, 3, 4 to achieve precise mirror control and ultimately correct the angular offset of the light spot. Since the auxiliary light and the writing light have different wavelengths, a beam splitter is used to separate photons from the auxiliary light, thereby enabling real-time power feedback monitoring and improving photon reception efficiency.

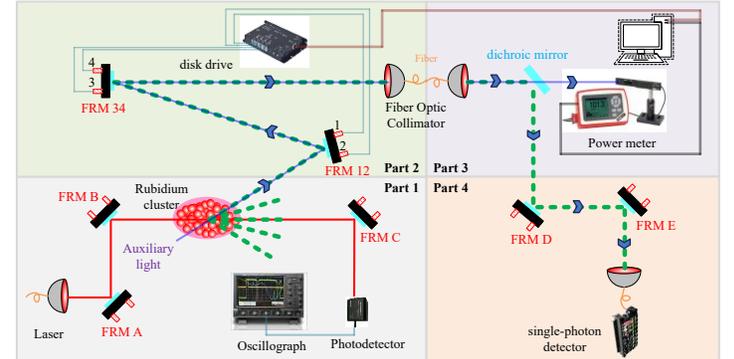

**Fig. 4.** Diagram of the experimental setup

The actuator generates small displacements using the difference between static and dynamic friction. The internal piezoelectric ceramic alternates between slow and fast electrical signals to drive relative sliding of the gripper structure, pushing the screw to achieve step rotation. The actuator receives commands from the host computer via a USB interface (e.g., "1-channel actuator moves 100 steps in the positive direction"), which are parsed into piezoelectric drive signal timing and voltage parameters to drive the actuator in "stepping mode". For a single actuator, the step size can remain constant under a fixed load. In this experiment, a single actuator step corresponds to a mirror deflection accuracy of 0.35 arcseconds. When the mirror was positioned 8.5 cm from the fiber collimator, the center-of-mass position changes were measured for different actuator step counts, recorded as center-of-mass positions X and Y. As shown in Fig. 5, each movement involved 10 steps, with movement step sizes of 10, 100, 200, 300, 400, 500, and their reverse movements. It can be observed that forward and reverse mirror movements are affected by hysteresis and nonlinear factors, resulting in differences in actuator forward and reverse step sizes for the same number of movements. Thus, positional accuracy effects caused by step size differences must be considered in subsequent adaptive compensation adjustments.

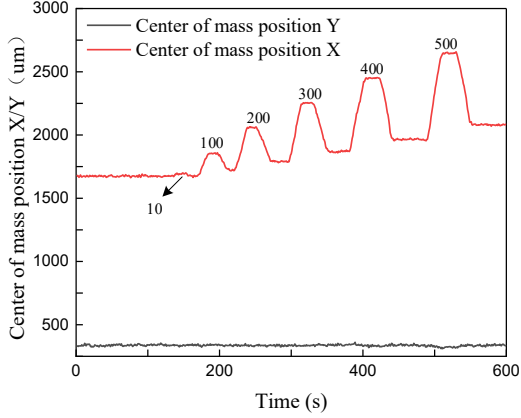

**Fig. 5.** Calibrated measurements of spot center-of-mass position during reflector deflection

*B. Control algorithms in feedback systems*

The single-peak symmetry and exponential decay property of the Gaussian mode field make the local optimization behavior of the hill-climbing algorithm naturally converge to the global optimal solution. The algorithm starts from the initial position (SetZeroPosition to zero) and can efficiently converge to the maximum power point at the center of the spot through dynamic step size and power gain evaluation $\Delta P = P_{new} - P_{old}$. The algorithm is based on the power feedback system. The feedback system utilizes a power feedback-based hill-climbing algorithm to perform the following steps:

1) Initialization phase: The system performs the first move in predefined steps (initial value of 100 or 10) starting from the initial position of the motor (zeroed by SetZeroPosition) and obtains the reference power value $P_0$ by means of an external power measurement file (Power_Val.txt);

2) Objective function evaluation: after each move, read the new power value $P_{new}$, and compare it with the previous power value $P_{old}$. Define the objective function gain as $\Delta P = P_{new} - P_{old}$. If $\Delta P > 0$, the power is boosted; otherwise, it indicates that the local optimum has been approached or exceeded.

3) Search direction decision: if $\Delta P > 0$, continue to move along the current direction (forward or reverse) and update $P_{old} = P_{new}$; if $\Delta P \leq 0$, perform a reverse move (with a negative step size) and mark the end of the local search (move_done = true) to avoid crossing the optimal solution.

4) Adaptive step size optimization: the algorithm adjusts the step size according to the number of outer cycles (cycles): initially, a large step size (e.g., 100 or 10) is used for a larger adjustment of the reflector to bring the spot closer to the center of the fiber, followed by an initial fine-tuning of the reflector with a smaller step size (e.g., 50 or 5) for a more accurate angle of incidence, then a large step size is reapplied to enable it to explore the power change region quickly, followed by a gradual reduction of its step size (e.g., 100-50-10 or 10-5-1) to precisely converge to the local maximum. Region, followed by a gradual reduction of its step size (e.g., 100-50-10 or 10-5-1) to accurately converge to a local maximum.

5) Iterative convergence and termination: the above steps are performed iteratively until the termination condition is satisfied: the number of adjustments reaches the upper limit or there is no further power gain in the neighborhood (i.e., $\Delta P \leq 0$). The final position is considered the optimal solution, and the relevant parameters (time, position, power values) are recorded to a CSV file.

The algorithm flow is shown in Fig. 6. The motor control itself is open-loop but introduces an external feedback mechanism by reading power values to achieve indirect closed-loop optimization, guided by power values. The motor continues moving in the current direction as long as power increases; when power decreases, the actuator stops and adjusts in the opposite direction. This approach bypasses hysteresis's direct effect on positional accuracy and improves search efficiency while reducing invalid adjustments and overshoots caused by nonlinearities to some extent through phased search with dynamic step size adjustment.

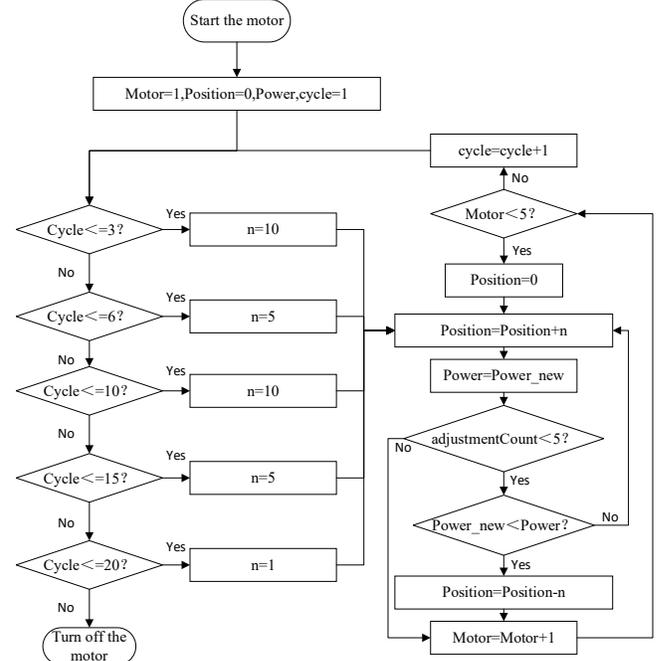

**Fig. 6.** Flowchart of power feedback-based hill-climbing algorithm

IV. EXPERIMENTAL VALIDATION AND RESULT ANALYSIS

*A. Experimental validation*

To verify the effectiveness of the designed cascaded automatic control device and its control algorithm, the experimentally constructed optical path schematic is shown in Fig. 7.

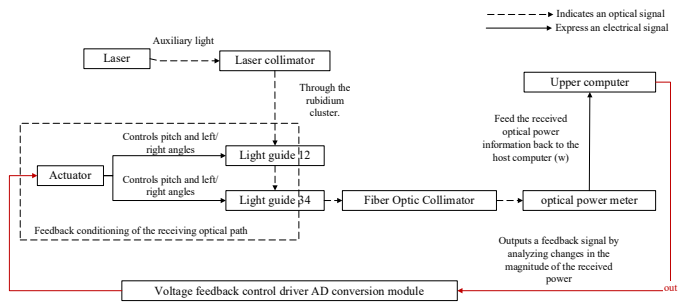

**Fig. 7.** Experimental schematic and system physical diagram

In the experiment, the SMF laser outputs 16.2 *mW* of optical power at the focal plane, with an optical power sampling frequency of 5 Hz. The experimental procedure is as follows: The laser is connected to a laser collimator via an SMF, and the collimated parallel light is reflected by light-guiding mirrors 12 and 34 before being received by the same collimator. The collimator position is then adjusted to ensure the beam remains parallel after passing through it. A multimode fiber is connected, and the optical path is roughly aligned to initially improve optical path quality. When the multimode fiber receiving power reaches approximately 90%, it is replaced with an SMF, followed by manual fine-tuning to maximize the SMF receiving power. The receiving efficiency is perturbed to reduce the received optical power to a certain value. Finally, the automatic control algorithm is activated to transmit the power value to the host computer, which controls the actuator's step count and step size based on the feedback power value and control algorithm to regulate the left-right tilting of mirrors 12 and 34. The algorithm's ability to restore the received optical power to the maximum value is observed, i.e., calibrating the angular offset at the receiving fiber.

*B. Result analysis*

The algorithm can be used for spatial light-multimode fiber coarse alignment, with step sizes of 100, 50, and 10 steps. The algorithm operation results are shown in Fig. 8.

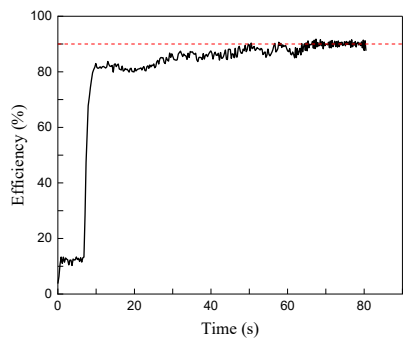

**Fig. 8.** Multimode fiber coupling efficiency curve

In a multimode fiber, the beam can propagate through multiple modes, so any small angular calibration rapidly improves coupling efficiency. As the number of iterations increases and the laser output stabilizes, the multimode fiber receiving efficiency improves rapidly, reaching 80% within 15 s. Finally, through fine calibration, rapid exploration of the power change region, and re-fine calibration, the receiving efficiency is improved to 90% within 80 s.

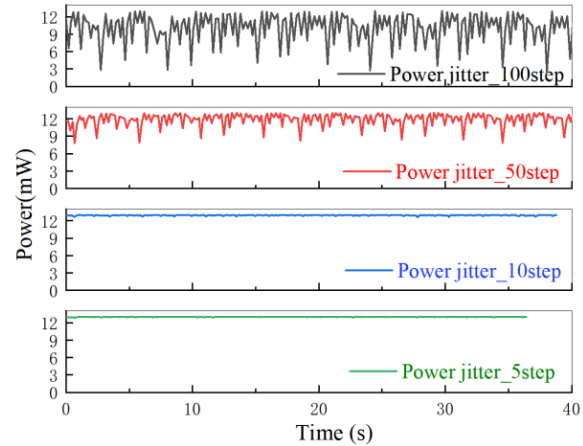

(a)

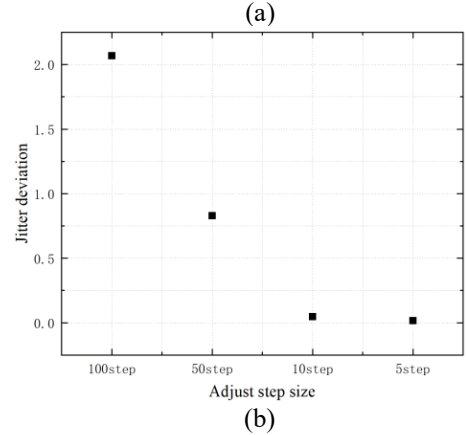

(b)

**Fig. 9.** Comparison of the degree of power jitter in the stabilization phase for different step sizes. (a) The distribution of power jitter with varying adjustment step sizes over time; (b) The distribution of jitter deviation as it varies with compensation.

When replacing the multimode fiber with a single-mode fiber (SMF), the latter exhibits higher sensitivity to angular deflection. If the same adjustment step size used for multimode fiber is retained during angle correction, significant power fluctuations will occur in the alignment stabilization phase, as illustrated in Figure 9(a). Here, the black, red, blue, and green lines represent the received power variations during the power stabilization period when using fixed deflection step sizes of 100, 50, 10, and 5, respectively. Experimental results indicate that an excessively large step size causes the system to frequently overshoot the optimal coupling position and enter the overshoot region, thereby reducing efficiency (Figure 9(b)). Notably, disturbance deviation decreases as the adjustment step size is reduced.

To mitigate overshoot during SMF alignment, a smaller movement step size should be employed in the algorithm convergence process. This study adopts a multi-step strategy: an initial large step size of 10, a medium step size of 5, and a minimum fine-tuning step size of 1. The SMF coupling efficiency curve is presented in Fig.10. In the experiment, after attenuating the power to the maximum stable position (specifically, power attenuation stabilizes without fluctuation around a fixed value rather than oscillating), the reflector position is continuously fine-tuned as power jitter diminishes, ultimately boosting the receiving efficiency to over 70%.

Experimental results confirm that the SMF receiving efficiency correction curve follows a "slow-fast-slow" trend, consistent with the Gaussian function attenuation law. The coupling efficiency to over 70% within 20 seconds and entering the most efficient and stable transmission state after 75 seconds.

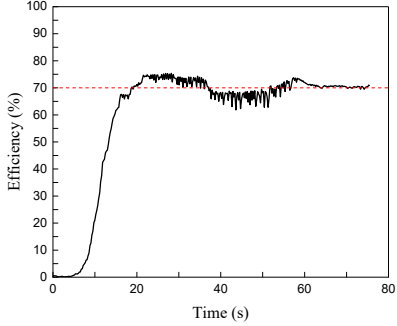

**Fig. 10.** Single-mode fiber coupling efficiency curve

The data fitting results are shown in Fig. 11. The theoretical curve is based on the exponential decay law of coupling efficiency with angular offset in the presence of intrinsic loss. The fitted data invert the angular offset using the theoretical formula $T = T_B \times e^{\left(-\frac{\pi\omega\theta}{\lambda}\right)^2}$, where base efficiency $T_B = 0.8$, $beam\ waist\ \omega = 1.625e-3$, wavelength $\lambda = 780e-9$, and the measured data point distribution characteristics support the effectiveness of angular calibration. Data points in the large angular deviation interval ($\theta > 0.2\ mard$) are sparse, corresponding to the calibration's coarse adjustment stage, where the system quickly covers large errors with large step sizes. Data points in the small angular deviation interval ($\theta \leq 0.1\ mard$) are dense, corresponding to the fine adjustment stage, where the system achieves small-angle fine convergence. The actual angular deviation continues to decrease and stabilizes within a very small range, reflecting the calibration process's deviation correction capability.

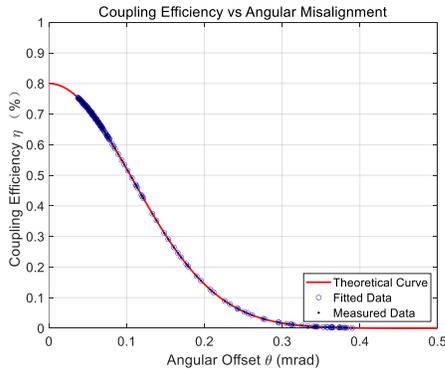

**Fig. 11.** Fitted curve of single-mode fiber coupling efficiency

## V. Conclusion

To address the unique challenges of spatial-fiber coupling in multimode quantum networks, this study proposes a cascaded adaptive feedback control scheme. By simulating the average light field of multiple modes via an auxiliary optical path and integrating a power-feedback-based hill-climbing algorithm to dynamically adjust mirrors, the scheme demonstrates three core advantages: efficient response, stable transmission, and precise control. It breaks the limitation of traditional single-mode correction that "iterates based on the number of modes" to achieve parallel correction of multiple modes—boosting the coupling efficiency to over 70% within 20 seconds and entering the most efficient and stable transmission state after 75 seconds. Moreover, it maintains a stable coupling efficiency of 70.74% (close to the intrinsic baseline efficiency in multimode scenarios) with single-node loss controlled within 30%. Meanwhile, through adaptive step-size optimization and piezoelectric actuation, the angular offset of the light spot is stabilized within 0.1 arcseconds, effectively avoiding mode crosstalk and suppressing quantum state decoherence, thus providing an efficient solution to the coupling challenges of multimode quantum networks.

The study establishes solid support in both theory and experiments. Theoretically, based on the mode-field matching principle, a spatial laser-single-mode fiber (SMF) angular offset model is constructed, which reveals that the coupling efficiency curve follows a Gaussian function decay law with the linear variation of angular offset. This clarifies the quantitative relationship between angular offset and coupling performance, laying a foundation for device design and algorithm optimization. Experimentally, a cascaded automatic control device is built; by adjusting the receiving efficiency of the auxiliary optical path to optimize the photon receiving path, it achieves rapid self-alignment of SMFs — improving the receiving efficiency from below 1% to 70.74% within 75 seconds. Furthermore, the staged characteristics of efficiency changes in the experiment are fully consistent with the theoretical law and the optimization process of mode-field matching, which fully verifies the reliability of the theoretical model and experimental system.

This research also holds significant application value and scalability potential. It not only overcomes the coupling bottleneck of multimode quantum networks and establishes a technical system featuring "rapid response, high efficiency and stability, and precise control" but also effectively suppresses angular jitter to stabilize the atom-photon interface (meeting the core requirement of quantum repeaters). In addition, the adaptive step-size design of the hill-climbing algorithm enables the device to be compatible with both multimode and SMF coupling, supporting the cascaded node architecture of large-scale networks. In conclusion, this study provides key technical support for efficient and stable spatial light transmission in scalable quantum networks and exhibits important application prospects in quantum entanglement source preparation and free-space coupling systems.